\documentclass{emulateapj}

\usepackage{apjfonts}

\newcommand{\etal}{et al.~}
\def\gsim{\lower 2pt \hbox{$\, \buildrel {\scriptstyle >}\over
{\scriptstyle \sim}\,$}}
\def\lsim{\lower 2pt \hbox{$\, \buildrel {\scriptstyle <}\over
{\scriptstyle \sim}\,$}}

\def\HI{H~{\scriptsize I}}

\def\ovi{O~{\scriptsize VI}}
\def\ovii{O~{\scriptsize VII}}
\def\oviii{O~{\scriptsize VIII}}

\def\cii{C~{\scriptsize II}}
\def\ciii{C~{\scriptsize III}}

\def\nii{N~{\scriptsize II}}

\def\neix{Ne~{\scriptsize IX}}

\shortauthors{Yao \& Wang}
\shorttitle{Galactic Diffuse Hot Gas Toward Mrk~421} 

\begin{document}
\slugcomment{\em Accepted for publication in the  Astrophysical Journal}

\title{The Non-isothermality and Extent of Galactic Diffuse Hot Gas
Toward Mrk~421}

\author{Yangsen Yao\altaffilmark{1} and Q. Daniel. Wang\altaffilmark{2}}
\altaffiltext{1}{Massachusetts Institute of Technology (MIT) Kavli Institute for Astrophysics and Space Research, 70 Vassar Street, Cambridge, MA 02139; yaoys@space.mit.edu}
\altaffiltext{2}{Department of Astronomy, University of Massachusetts, 
  Amherst, MA 01003; wqd@astro.umass.edu}

\begin{abstract}

Diffuse hot gas can be traced effectively by its X-ray absorption and 
emission. We present a joint-analysis of these tracers to characterize the spatial and 
temperature distributions of the 
Galactic hot gas along the sight-line toward the nearby bright active 
galactic nucleus Mrk 421. We also complement this analysis with
far-UV \ovi\ absorption observations. We find that the observed absorption 
line strengths of \ovii\ and \oviii\ are {\sl inconsistent} with
the diffuse background emission line ratio of the same ions, if 
the gas is assumed to be isothermal in a collisional ionization equilibrium 
state. But all these lines as well as the diffuse 3/4-keV broad-band background
intensity in the field can be fitted with a plasma with a
power law temperature distribution. We show that this distribution
can be derived from a hot gaseous disk model with the gas temperature 
and density decreasing exponentially with the vertical
distance from the Galactic plane. The joint fit gives the 
exponential scale heights as $\sim 1.0$ kpc and 
$1.6$ kpc and the middle plane values
as $2.8\times10^6$ K and $2.4 \times 10^{-3}\ {\rm cm^{-3}}$
for the temperature and density, respectively. 
These values are consistent with those inferred from X-ray observations of 
nearby edge-on galaxies similar to our own. 

\end{abstract}

\keywords{Galaxy: disk --- Galaxy: structure --- X-rays: ISM --- X-rays: individual (Mrk 421)}

\section{Introduction}

X-ray absorption lines of \ovii, \oviii, and/or \neix\ at $z\sim0$
have been detected unambiguously toward several
bright AGNs, but the nature of these highly-ionized species remains
uncertain. Because of the limited spectral resolution
of the current grating instruments on board the
{\sl Chandra} and  {\sl XMM-Newton X-ray observatories} 
 (FWHM $\sim 50$ m\AA, corresponding to a velocity dispersion of 
750 km~s$^{-1}$ at 20~\AA), however, the
line width and centroid position do not directly constrain the absorbing gas
to be any significantly better than a few Mpc.
Some authors have attributed these $z\sim0$
absorptions to the so-called warm hot intergalactic medium (WHIM) 
in the Local Group or
even beyond (e.g., Nicastro \etal 2002; Rasmussen \etal 2003;
Williams \etal 2005).
In this interpretation, the hot phase medium could contain 
large amounts of baryonic matter, accounting for the bulk of 
the ``missing baryons'',
as predicted in numerical simulations of large-scale structure formations
(e.g., Cen \& Ostriker 1999; Dav\'e \etal 2001).

But, the $z\sim0$ \ovii, \oviii, and/or \neix\ absorption lines 
have also been observed in X-ray spectra of the LMC X-3 at distance of
50 kpc (Wang et al. 2005) and many Galactic sources
(e.g., Futamoto \etal 2004; Yao \& Wang 2005; Juett \etal 2006). 
The strength of the absorption
toward such a source, depending its Galactic coordinates and pathlength, 
can be substantially greater than those toward the AGNs. A crude
characterization of the global 
distribution of the absorbing hot gas gives an effective
scale length of several kpc, suggesting that much of the 
observed absorptions originates in the Galaxy \citep{yao05}.
The recent surveys of AGN spectra with the {\sl Chandra} and {\sl XMM-Newton} 
archive data show that the \ovii\ absorption line is present in all 
high quality spectra \citep{mck04, fang06}, indicating a covering fraction
of \ovii\ absorbing gas $\gsim63\%$ in the sky. This high covering fraction, 
together with an estimation of the total mass of the 
absorbing gas and with a comparison between emission and absorption 
measurements, further supports the scenario that the X-ray absorbers
are primarily associated with the Galaxy \citep{fang06}. There is really
little doubt as to the presence of the Galactic contribution to the 
X-ray absorption. The question is how much. 

While the line absorption depends on the ionic column density and the 
dispersion velocity of the intervening hot gas, the emission 
is sensitive to the number density and temperature of the gas.
Therefore a combination of absorption and emission measurements 
naturally provides
a constraint to the scale length of the absorbing/emitting plasma. 
\citet{ras03} and \citet{fut04} have indeed compared the observed 
\ovii\ absorption with the
background \ovii\ emission line intensity \citep{mcc02} to estimate the scale 
length of the hot gas around the Galaxy. Similar estimates
are also made by \citet{fang06}. But, so far such estimates all depend
on various {\sl assumptions} such as absorption in the linear regime 
and isothermality of the hot gas, which may not be valid 
(see \S\ref{sec:absline} and \S4 for further discussion).

Here we report the first systematic joint-analysis of X-ray absorption 
and emission data, complemented by a far-UV OVI absorption measurement,
in the direction of Mrk~421. The joint-analysis allows for tests of 
the assumptions and provide tight constraints on our spectral models.
Our goals are to characterize both the scale length 
and global thermal property of the $z \sim 0$ absorbing/emitting hot gas
toward Mrk~421.

Throughout the paper, the  statistical errors are quoted at 
the 90\% confidence level, unless being pointed out specifically. We also 
adopt the solar abundance given by \citet{and89} for oxygen\footnote{
  We note that \citet{asp05} proposed
  a revision of the commonly used solar abundances of \citet{and89},
  but this revision is still under debate \citep{ant06}.
  We therefore still use the old values.
}.

\section{Observations and Data Reduction \label{sec:data}}

\subsection{{\sl Chandra} X-ray Absorption Data} \label{sec:chandra}

Mrk~421 is one of the brightest quasars at $z=0.03$ 
($l, b= 179\fdg83, 65\fdg03$), and its X-ray flux varies
by up to $\sim50$ times between quiescent to burst states. 
Table~\ref{tab:observation} lists the existing nine {\sl Chandra} 
observations of Mrk~421, covering various flux levels of the source. 
Two observations used the High
Energy Transmission Grating (HETG; Canizares \etal 2005) with the 
Advanced CCD Imaging Spectrometer (ACIS-S; Garmire \etal 2003) and the other
seven used the Low Energy Transmission Grating (LETG) with either
the ACIS-S (five times) or the High Resolution Camera (HRC-S; Murray \& Cappell 1986).
The LETG provides $\sim30-40~{\rm cm^2}$ effective area around 
\ovii\ and \oviii\ K$\alpha$ wavelength with a modest resolution of
$\sim 750\ {\rm km\ s^{-1}}$.
\citet{nic05} and \citet{wil05} have reported three 
(ObsID 1715, 4148, and 4149) of these observations.
In this work, we combine six long exposures of 
these observations to enhance the 
counting statistics (by $\sim30\%$); the other three observations, which we do
not use, have short exposures, contributing only $<5\%$ to the 
total recorded counts around the wavelength of \ovii\ K$\alpha$ (the 
most significant absorption line). 
Furthermore, the LETG-ACIS combination suffers no grating order over-lapping complexity as  
the LETG-HRC does, allowing for a sensitive cross-checking of
the measurements between the ACIS-S and HRC-S observations 
(see below and \S4).

\begin{deluxetable}{llccc}[!h]
\tablewidth{0pt}
\tablecaption{{\sl Chandra} Observations\label{tab:observation}}
\tablehead{
       &           &                  & Exposure &\\
 ObsID & Obs. Date & Grating/Detector & (ks)     & Used}
\startdata
457  & 1999 Nov 5  & HETG-ACIS & 19.83 & No \\
1714 & 2000 May 29 & HETG-ACIS & 19.84 & No \\
1715 & 2000 May 29 & LETG-HRC  & 10.89 & No \\  
4148 & 2002 Oct 26 & LETG-ACIS & 96.84 & Yes\\
4149 & 2003 Jul 1  & LETG-HRC  & 99.98 & Yes\\
5171 & 2004 Jul 3  & LETG-ACIS & 67.15 & Yes\\
5318 & 2004 May 6  & LETG-ACIS & 30.16 & Yes\\
5331 & 2004 Jul 12 & LETG-ACIS & 69.50 & Yes\\
5332 & 2004 Jul 14 & LETG-ACIS & 67.06 & Yes\\
\hline                   
\enddata                 
\end{deluxetable}

We follow the same procedure as described in \citet{yao06} to
re-process the individual observations. For 
the ACIS-S observations, we extract the first grating order spectra and 
calculate the corresponding response matrix files 
(RMFs) and auxiliary response functions (ARFs). 
For the HRC-S observation, we extract the total grating spectrum and calculate
the RMFs and ARFs from the first to the sixth grating orders, and then
use IDL routines from PINTofALE package\footnote{http://hea-www.harvard.edu/PINTofALE}
to combine all the RMFs and ARFs
pairs to form an order-integrated response file accounting for the higher order
overlapping in the spectral fitting. To improve the counting statistics, 
all the spectra, after a consistency check, are then co-added and the 
corresponding ARFs and RMFs are
weight-averaged according to the exposure and 
global continuum intensity obtained from a model fitting 
to each spectrum. In the final co-added spectrum, we obtain about 
2700 and 1600 counts per channel with a channel width of 12.5 m\AA\ around 
wavelength of 18.5, and 21.5 \AA, where our interested absorption 
lines are located. 

Our global continuum modeling is conducted in the 2--22.5 \AA\ range. 
We adopt a power law model with a foreground cool gas absorption to 
characterize the overall spectral shape of Mkr~421.
 As in previous studies \citep{nic05, wang05}, 
in order to obtain an acceptable global fit, we include 11 broad 
($\sigma>1500\ {\rm km\ s^{-1}}$) Gaussian profiles to account for
spectral deviations from the adopted power law model and for
various calibration uncertainties in detector gap/node regions. 
The \neix, \oviii, \ovii~K$\alpha$, and \ovii~K$\beta$ absorption lines 
are clearly visible in their corresponding rest frame wavelength (we do not 
use other marginally significant $z\sim0$ absorption lines that have 
been reported by Williams \etal 2005 and by Nicastro \etal 2005). These
absorption lines are characterized with narrow Gaussian profiles in 
the continuum fit. 
We obtain a final spectral fit with $\chi^2/d.o.f=1987/1577$, which, 
given such high counting statistics, is reasonably acceptable and in 
particular is 
good enough for accounting for the higher order overlapping in the LETG-HRC
data over the whole 
wavelength range. The $\chi^2/d.o.f$ in the 12.5-14.5 \AA\, 18.2-20.0 \AA,
and 20.5-22.5 \AA\ ranges are 202/148, 180/148
and 202/147, respectively.

\subsection{Diffuse X-ray Emission Data \label{sec:sxb}}

We use the diffuse 3/4-keV band (R4+R5 bands) background intensity from 
the {\sl ROSAT All Sky Survey} (RASS; Snowden \etal 1997) to estimate the 
emission from the hot gas in the field of Mrk 421. The 
background at lower energies (i.e., 1/4-keV band) 
is subject to strong absorption 
by the cool interstellar medium and therefore 
traces mostly the hot gas within the Local Bubble. The background
in the 1.5-keV band is primarily point-like in origin. 
To avoid any significant contamination from the bright Mrk~421 due to 
the extended point spread function wing of the {\sl ROSAT}, 
we use an annulus between $0\fdg75$ and $1\fdg75$ radii
around the source to obtain a 3/4-keV background intensity 
estimate as $\sim1.06\times 10^{-4}\ {\rm counts\ s^{-1}\ armin^{-2}}$.
This background intensity contains a significant contribution from 
unresolved point-like sources.
The extragalactic point-like source contribution 
can be characterized as a power law function with photon index 
1.4 and a normalization
$10.9\ {\rm photons\ s^{-1}\ keV^{-1}\ cm^{-2}\ sr^{-1}}$ at 1 keV
\citep{hic06}, which is equivalent to 
$5.61\times 10^{-5}\ {\rm counts\ s^{-1}\ armin^{-2}}$ 
in the RASS 3/4-keV energy 
band (after considering the foreground cool gas absorption
$N_H^c = 1.43\times10^{20}\ \rm{cm}^{-2}$; Dickey \& Lockman 1990).
The Galactic component (due to stars) is
$\sim6 \times 10^{-6}\ {\rm counts\ s^{-1}\ armin^{-2}}$ at the 
Galactic latitude of 
Mrk~421 \citep{kun01}. Therefore, we estimate
the background intensity due to the Galactic
diffuse hot gas as $4.6\times 10^{-5}\ {\rm counts\ s^{-1}\ armin^{-2}}$.
Here we have neglected a potential contribution from charge exchange 
interactions of solar wind ions with the local neutral 
interstellar medium (ISM; e.g., Pepino \etal 2004).

The 3/4-keV background is predominantly due to the \ovii\ and \oviii\ 
K$\alpha$ line emission. The best existing spectroscopic measurement of
such line emission is from 
the micro-calorimeters flying on a sounding
rocket \citep{mcc02} over
a region of $\sim1$ sr, centered at 
$l, b = 90^\circ, 61^\circ$, which is at nearly the same Galactic latitude 
as the Mkr~421 direction and is certainly away from abnormal regions such as
the Galactic Bulge, the North Polar Spur, and the Cygnus Loop. 
The measured line intensities are $I_{\rm OVII}=4.8\pm0.8$ and 
$I_{\rm OVIII}=1.6\pm0.4\ {\rm photons\ s^{-1}\ cm^{-2}\ sr^{-1}}$ 
(1 $\sigma$ errors). We assume these intensities are typical for the
hot gas at high latitudes.
The effects of the foreground cool gas absorption are expected to be small
\footnote{The neutral hydrogen column density toward the rocket experimental 
direction is $1.9\times10^{20}~{\rm cm^{-2}}$. 
The absorption correction will increase the OVII and OVIII line intensities 
by 18\% and 12\% respectively; these changes are comparable to or less than
the corresponding 1$\sigma$ statistic uncertainties.}
and therefore have not been considered in this work.
We use these two line intensities, implemented in a two data point 
spectrum, to characterize the hot gas emission. 
This spectrum will be jointly fitted with the absorption data to constrain 
the emitting/absorbing gas properties, whereas the broadband  
RASS 3/4-keV intensity
will be utilized as a consistency check. We also fake a 
response file containing two delta functions for this spectrum, so this
emission-line-intensity spectrum can be directly compared to our emission
line model in the joint analysis.  

\subsection{{\sl FUSE} \ovi\ Absorption Data \label{sec:fuse}}

{\sl FUSE}\footnote{For detail description of {\sl FUSE}, please refer to
\citet{moo00}}
 observed Mrk~421 four times with 29 individual exposures and an 
actual integrated exposure of 84 ks in total, all configured with 
the target aperture LWRS ($30''\times30''$), providing
effective area of $\sim50\ {\rm cm^2}$ with spectral resolution of
$\sim20\ {\rm km\ s^{-1}}$ around the \ovi\ doublet (1032 and 1038 \AA).

We only use the data from the LiF1a segment. We calibrate the wavelength for each
exposure with CalFuse v3.0\footnote{http://fuse.pha.jhu.edu/analysis/calfuse.html}, and then
repeatedly cross-correlate the wavelength scale of different exposures and 
co-add them to form a final combined spectrum using Don Lindler's IDL Tools\footnote{http://fuse.pha.jhu.edu/analysis/fuse\_idl\_tools.html}.
In order to jointly-analyze the {\sl FUSE} and X-ray spectra together in 
software package XSPEC, we transfer the {\sl FUSE} FITS spectrum to a 
format that can be recognized in XSPEC. We adopt a Gaussian
line spread function with 20 ${\rm km\ s^{-1}}$ (FWHM) 
as the spectral response of the {\sl FUSE} spectrum.

We use a power law to approximate
the continuum and a Gaussian profile to characterize the \ovi\ absorption.
We find that one broad Gaussian component is adequate to describe the total
absorption. We obtain an EW of the \ovi\ line that is consistent with the
previously reported value (e.g., Wakker \etal 2003;
Williams \etal 2005). As the line is well-resolved, we measure
the  dispersion velocity and \ovi\ column density of the absorbing gas as 
$v_b$ 
\footnote{We use $v_b$ instead of the more generally used $b$ to denote the 
standard velocity dispersion to avoid the confusion with the Galactic latitude.}
=  90(84, 97) km~s$^{-1}$ and log[$N{\rm _{OVI}(cm^{-2})}$] 
= 14.44(14.41, 14.47). 

\section{Data Analysis and Results \label{sec:results}}

Our analysis includes the absorption and emission 
lines of oxygen only to avoid uncertainties in the relative elemental 
abundances of the hot gas.
We assume that the hot gas is in a collisional ionization equilibrium 
(CIE) state and that the X-ray absorption contains only one
velocity component. In reality, each observed line could 
consist of multiple velocity components. 
With the still limited spectral resolution of current X-ray data, we decide not to
investigate this complexity and leave it to the future when better 
data become available.
The absorption lines of various transients are jointly fitted with
our model {\sl absline}\footnote{
For a detail description
of the {\sl absline}  model and its application in multiple absorption 
line diagnostics, please refer to \citet{yao05, yao06} and \citet{wang05}.
},
implemented in the spectral analysis software {\sl XSPEC}. 

\subsection{Isothermal Model \label{sec:absline}}

\begin{figure}[h]
  \centerline{
    \includegraphics[width=0.45\textwidth]{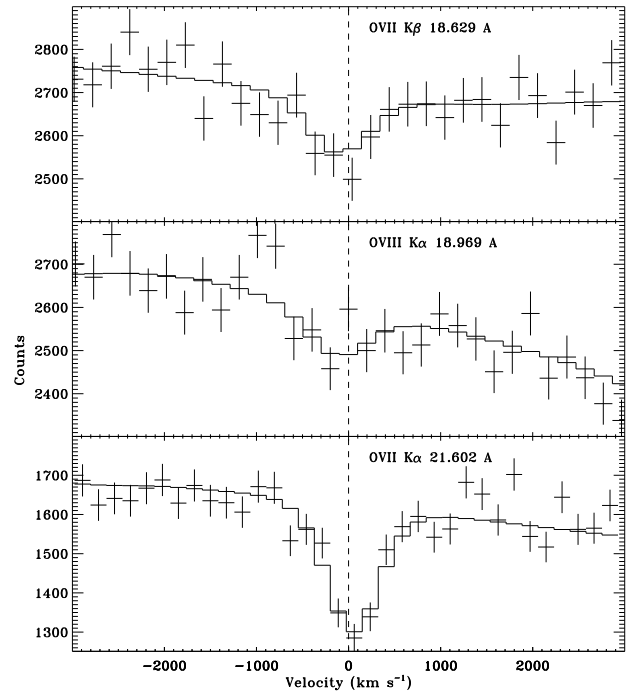}
  }
  \caption{The joint fit of the X-ray absorption lines with an
    isothermal {\sl absline} model.
    The bin size is 12.5 m\AA; the ARF and RMF
    have been applied to the model. 
    \label{fig:OVI}
  }
\end{figure}

We first conduct a joint analysis of the \ovii\ K$\alpha$, \ovii\ K$\beta$, 
 and \oviii\ K$\alpha$ X-ray absorption lines (Fig.~\ref{fig:OVI}).
The goodness of the spectral fit with the {\sl absline} model is comparable to 
that obtained with the Gaussian profiles (\S\ref{sec:chandra}).
From the fit, we obtain the dispersion velocity $v_b$, \ovii\ column density 
$N_{\rm OVII}$, and the temperature $T$ of the intervening gas. We can then 
infer the \ovi, \oviii, and hydrogen column densities 
($N_{\rm OVI}$, $N_{\rm OVIII}$, and $N_{\rm H}$ 
respectively) of the 
absorbing gas (Table~\ref{tab:abs}).

\begin{deluxetable*}{lcccccc}[!h]
\tabletypesize{\small}
\tablewidth{0pt}
\tablecaption{
  Joint analysis results   \label{tab:abs} }
\tablehead{
  & $v_b$ & log(T) & log($N_{\rm OVI}$) & log($N_{\rm OVII}$) & log($N_{\rm OVIII}$) & log($N_{\rm H}A_{\rm O}$)$^a$ \\
  & (km s$^{-1}$) & (K) & (cm$^{-2}$)    & (cm$^{-2}$)         & (cm$^{-2}$)          & (cm$^{-2}$) }
\startdata
\multicolumn{7}{l}{\underline{Isothermal Modeling}}\\
abs.  & 64(48, 104) & 6.16(6.10, 6.21) & 13.51(13.32, 13.70)$^b$ & 16.00(15.82, 16.17) & 15.18(14.78, 15.41) & 19.14(18.98, 19.30)\\
\multicolumn{7}{l}{\underline{Non-isothermal Modeling}}\\
abs. + emi. & 64(48, 103) & 6.44(6.36, 6.54)$^c$ & 14.76(14.53, 14.81)$^{b,d}$ & 15.99(15.76, 16.04)$^d$ & 15.24(15.01, 15.29)$^d$ & 19.29(19.06, 19.44)\\
X-ray + FUV$^e$ & 90(83, 96) & 6.44(6.37, 6.52)$^c$ & 14.44(14.38, 14.51)$^d$ & 15.86(15.80, 15.93)$^d$ & 15.23(15.17, 15.30)$^d$ & ~19.12(19.06, 19.19)
\enddata
\tablecomments{In our absorption line model, the transition oscillation 
  coefficient and the damping factor are adopted from \citet{ver96}
  for \ovii\ and \oviii, and from \citet{mor03} for \ovi.
  $^a$ $A_{\rm O}$ is oxygen abundance in unit of solar.
  $^b$ The model predicated values.
  $^c$ The logarithmic value of $T_{\rm 0}$ at the Galactic plane.
  $^d$ These uncertainty ranges can not be directly calculated from the 
  joint analysis, and the error ranges listed here are scaled to have the 
  same ranges as that of $N_{\rm H}$.
  $^e$ Assuming that all \ovi\ absorptions are associated with 
  \ovii-bearing gas. See text for detail.
}
\end{deluxetable*}

\begin{figure}[h]
  \centerline{
      \includegraphics[width=0.45\textwidth]{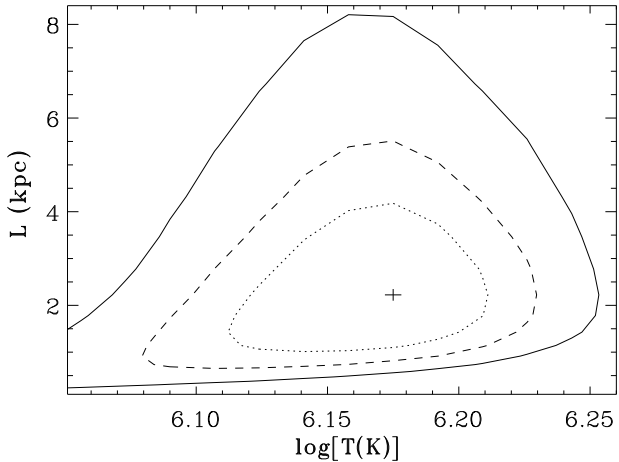}
  }
  \caption{The 68\%, 90\%, and 99\% confidence contours of 
    effective scale length ($L$) versus the temperature ($T$) of the 
    emitting/absorbing gas at the CIE state, which is obtained by
    jointly-fitting the \ovii\ (without \oviii) emission line intensity 
    with the observed X-ray \ovii\ and \oviii\ absorption lines.
    \label{fig:L_T0_ST} }
\end{figure}

\begin{figure}[h]
  \centerline{
   \includegraphics[width=0.45\textwidth]{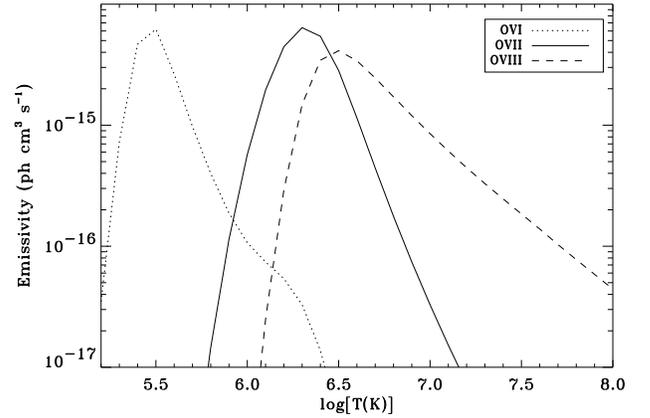}
  }
  \caption{The \ovi, \ovii, and \oviii\ line emissivities as a function of
   the gas temperature. The value for \ovi\ has been scaled down by a
    factor of 1000 for ease of demonstration.
    \label{fig:oxygen}
  }
\end{figure}

We further jointly analyze the \ovii\ emission line intensity 
(without including the \oviii\ line; \S\ref{sec:sxb}) 
with the X-ray absorption lines to estimate the scale length of the 
emitting/absorbing gas.
The oxygen emission line intensity is
\begin{equation} \label{equ:eline}
  I [{\rm ph\ cm^{-2}\ s^{-1}\ sr^{-1}}] 
  = \frac{A_{\rm O}}{4\pi} \int_0^L \Lambda(T) n_e n dl,
\end{equation}
where $A_{\rm O}$ is the oxygen abundance (in unit of the 
solar value and assumed to be 
1 in the paper), 
$\Lambda(T)$ is the line emissivity\footnote{
  The definition of $\Lambda(T)$ is adopted from 
  http://cxc.harvard.edu/atomdb/physics/physics\_units/physics\_units.html.
},
and $L$ is the effective path length of the gas.
Assuming that the hydrogen number density can be characterized  as 
\begin{equation} \label{equ:expHy}
  n  =  n_0 e^{-z/(h_{\rm n}\xi)},
\end{equation}
where  $z$ is the vertical distance away from the Galactic plane, 
$n_{\rm 0}$ and $h_{\rm n}$ are the mid-plane value and the scale height,
and $\xi$ is the volume filling factor that is assumed to be 1 in the paper, 
we can integrate Eq.~\ref{equ:eline} along a sight-line
with a Galactic latitude $b$ to get
\begin{equation} \label{equ:elineST}
  I   = \frac{1.2 A_{\rm O} N_H^2 \Lambda(T)}{8\pi L}.
\end{equation}
Here the factor 1.2 accounts for the helium 
contribution to the electron density, $L=h_{\rm n}\xi/\sin b$,
$N_{\rm H}=n_{\rm 0} h_{\rm n}\xi/\sin b$.
We construct a simple emission line 
model based on Eq.~\ref{equ:elineST}, and use it to model the
emission line intensity spectrum we have constructed (\S\ref{sec:sxb}). 
In the joint analysis, $N_{\rm H}$ and $T$ are the common fitting 
parameters in both the absorption and emission line models, whereas $L$ only
appears in the emission model. The obtained $N_{\rm H}$ and $T$ values are 
nearly identical to those from fitting the absorption lines alone. 
In addition, we constrain 
$L$ to be $\sim 2(0.5, 4)$ kpc. Fig.~\ref{fig:L_T0_ST} presents the
68\%, 90\% and 99\% confidence contours of $L$ versus $T$. 
However, this isothermal modeling predicts an emission line intensity
ratio of $I_{\rm OVII}/I_{\rm OVIII}\gsim20$, in contrast to
the observed value of 3 (\S\ref{sec:sxb}). 
In the isothermal scenario, this ratio depends only on $T$
(Eq.\ref{equ:elineST}; Fig.~\ref{fig:oxygen}).  
To reproduce this ratio, the temperature needs to be $\sim10^{6.33}$ K, 
which is significantly higher than 
what is inferred from the isothermal model (at  
$>99\%$ confidence; Fig.~\ref{fig:L_T0_ST}).
This inconsistency is apparently due to the over-simplification of
the isothermality for the hot gas.

\subsection{Non-isothermal Model \label{sec:NIM}}

We now consider a simple non-isothermal
hot gaseous disk model to interpret the absorption and emission data.
This model is motivated by the thick disk-like morphology of 
diffuse hot gas observed in nearby normal edge-on galaxies 
and inferred from our statistical sight-line dependence analysis of 
the \neix\ absorption line (Yao \& Wang 2005; T\"ullmann \etal 2006; 
see \S4 for the discussion). In addition to the characterization
of the density distribution as Eq.~\ref{equ:expHy},
the temperature distributions of the hot gas
in the disk is assumed to be 
\begin{equation} \label{equ:expT}
  T  =  T_{\rm 0} e^{-z/(h_{\rm T}\xi)},
\end{equation}
where the mid-plane value $T_{\rm 0}$ and the scale height $h_{\rm T}$, along 
with $h_{\rm n}$ and $n_{\rm 0}$ in Eq.~\ref{equ:expHy}, are to be determined.
This temperature distribution is particularly motivated by the apparently
lower temperatures of the absorbing gas toward LMC~X--3 and Mrk~421 than
those toward Galactic sources \citep{yao05, wang05}.

From Eqs.~\ref{equ:expHy} and \ref{equ:expT}, we can derive
\begin{equation} 
n=n_{\rm 0}(T/T_{\rm 0})^\gamma,
\end{equation}
where $\gamma=h_{\rm T}/h_{\rm n}$.
Therefore, the differential 
hydrogen column density distribution is also a power law:
\begin{equation} \label{equ:PL}
  \begin{array}{rl}
    dN_H = & n dL \\
       = & \frac{N_{\rm H}\gamma}{T_{\rm 0}} (T/T_{\rm 0})^{\gamma-1} dT.
    \end{array}
\end{equation}
The corresponding ionic column density can  be expressed as
\begin{equation} \label{equ:dNi}
  N_{\rm i} = \frac{N_{\rm H} \gamma A_{\rm e} }{T_{\rm 0}}
  \int^{T_{\rm 0}}_{T_{\rm min}} 
  \left( \frac{T}{T_{\rm 0}} \right)^{\gamma-1} f_{\rm i}(T)dT,
\end{equation}
where 
$A_{\rm e}$ and $f_{\rm i}(T)$ are the element abundance 
and ionization fraction, and $T_{\rm min}$ is the minimum temperature
considered in this work (see below). 

Similarly, the surface emission intensity is 
(Eq.~\ref{equ:eline})
\begin{equation} \label{equ:int}
    I  =  \frac{A_e}{4\pi}\int^{T_{\rm 0}}_{T_{\rm min}}
	\Lambda(T) \frac{dEM}{dT}dT
\end{equation}
where  
\begin{equation} \label{equ:EMpl}
  \frac{dEM}{dT} = \frac{1.2 N_{\rm H}^2 \gamma}{T_{\rm 0 }L}
  \left( \frac{T}{T_{\rm 0}} \right)^{2\gamma-1}.
\end{equation}
The intensity can then be
calculated for an individual emission line or a broad band with the
appropriately chosen $\Lambda$. 

We modify our {\it absline} model to accommodate the power law temperature 
dependence of the absorbing hot gas column density (Eq.~\ref{equ:dNi})
and also revise the emission line model constructed in \S\ref{sec:absline}
according to Eq.~\ref{equ:int}. We then use these models to jointly fit
the X-ray absorption and emission data (including both \ovii\ and \oviii\
emission lines). We consider the hot gas in the temperature range 
from $T_{\rm min}=10^{5.0}$ K to $T_{\rm 0}$; gas at lower ($<10^5$ K)
temperatures contributes little to the \ovi, \ovii, and \oviii\ line 
absorption/emission.
The absorption/emission line joint fit with this non-isothermal model is
as good as 
the one to the absorption lines alone (\S\ref{sec:absline}).
We obtain log[$T_0({\rm K})$] = 6.44(6.36, 6,54), $L=1.8(0.5, 4.5)$ kpc,
and $\gamma\lsim1.4$.
The fitted $N_{\rm H}$ and $v_b$ values, as included in 
Table~\ref{tab:abs}, are consistent with those obtained in the isothermal
modeling (\S~\ref{sec:absline}).
But the model predicated \ovii/\oviii\ emission line ratio, 
3.7(2.2, 6.7), is now consistent with the observed value.

\begin{figure*}[h]
  \centerline{
      \includegraphics[width=0.95\textwidth]{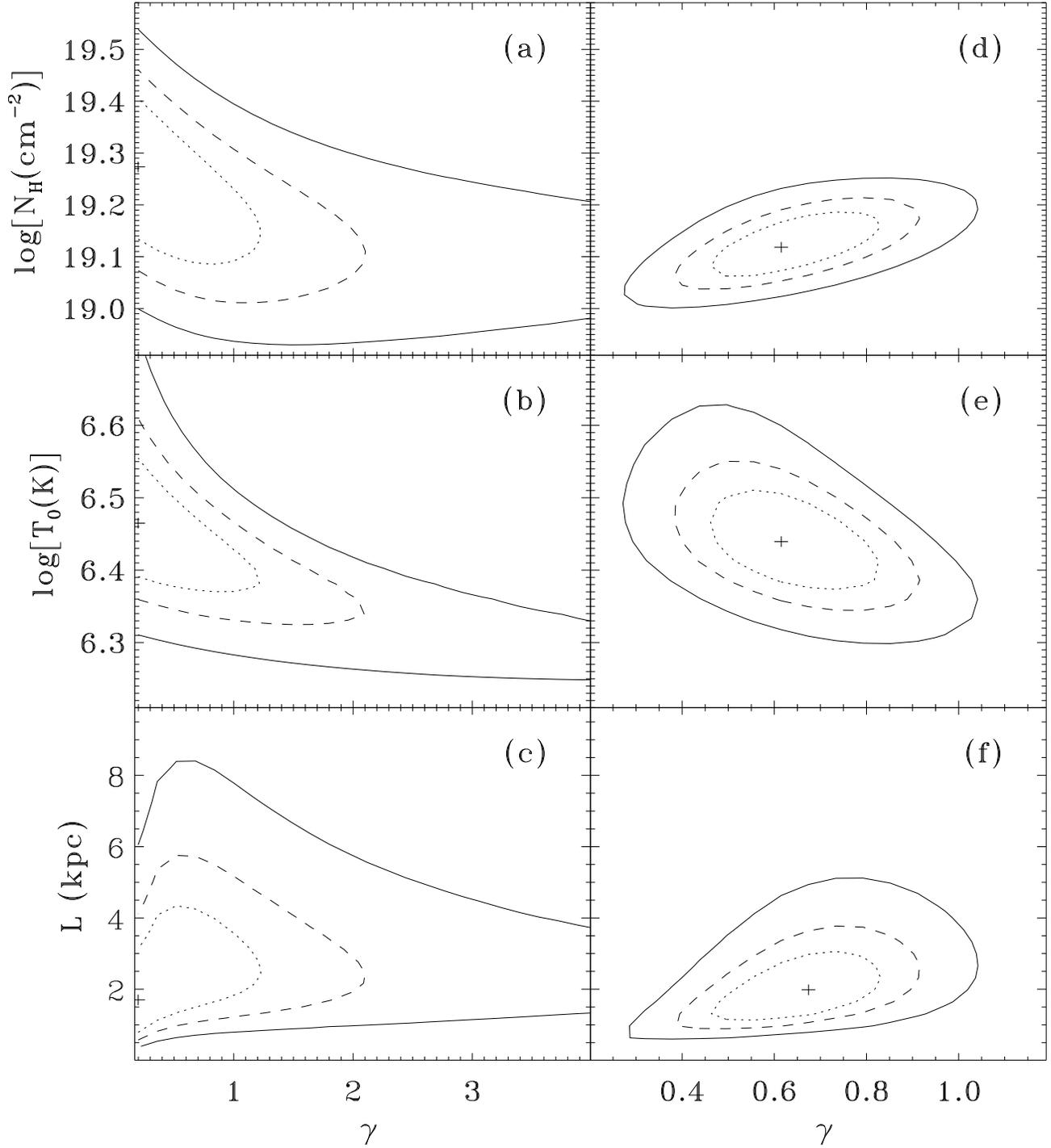}
  }
  \caption{\small{The 68\%, 90\%, and 99\% confidence contours of the 
    non-isothermal model parameters. They are obtained from 
    the joint analysis the X-ray absorption/emission data with 
    (right panels) and without (left panels) the inclusion of the 
    far-UV \ovi\ absorption line. See text for details.
    \label{fig:Gamma_Tmax}}
  }
\end{figure*}

It is constructive to specify how the individual parameters are constrained,
although they are generally correlated in the spectral fitting. The 
relative \ovii\ K$\alpha$ to K$\beta$ absorption line strength provides 
constraints 
mainly on $v_b$, $N_{\rm OVII}$/$N_{\rm OVIII}$ on $\gamma$, $N_{\rm OVII}$ 
on $N_{\rm H}$ (via the assumed $A_{\rm O}$),
the emission line intensity ratio $I_{\rm OVII}/I_{\rm OVIII}$ 
on $T_{\rm 0}$, and the ratio of $I_{\rm OVII}$ to $N_{\rm OVII}$ on $L$.
Fig.~\ref{fig:Gamma_Tmax}(a)-(c)
presents the confidence contours of $N_{\rm H}$, $T_{\rm 0}$, and $L$ versus 
$\gamma$. It is clear 
that although the lower limit to $\gamma$ is poorly determined here,
the $L$ is well constrained to be $<10$ kpc at the 99\% confident level.

While the X-ray data do not give a tight constraint on $\gamma$, 
the inclusion of the {\sl FUSE} \ovi\ absorption observation in our fit can be
helpful. In fact, the above non-isothermal modeling with the X-ray data alone
over-predicts the \ovi\ column density by a factor of $\sim2$, 
judged by the best-fitted non-isothermal model 
(Table~\ref{tab:abs}, \S\ref{sec:fuse}). If the \ovi-bearing 
gas is entirely located in the hot gaseous disk, as prescribed for
the X-ray absorption and emission,
we can then obtain $\gamma=0.6(0.4, 0.9)$, $L=1.8(1.0, 3.2)$ kpc, while the
other parameters are included in Table~\ref{tab:abs} for comparison.
The disk model parameters are log$[T_{\rm 0}({\rm K})]=6.44(6.37, 6.52)$,
$n_{\rm 0}=2.4(1.6, 3.7)\times10^{-3}\ {\rm cm^{-3}}$, 
$h_{\rm n}=1.6(0.9, 2.9)/\xi$ kpc, and 
$h_{\rm T} = 1.0(0.6, 1.8)/\xi$ kpc.
The confidence contours for $N_{\rm H}$, $T_{\rm 0}$, and $L$ versus 
$\gamma$ are presented in Fig.~\ref{fig:Gamma_Tmax} (d)-(f). Compared to 
Fig.~\ref{fig:Gamma_Tmax} (a)-(c), 
the tightening is partly due to the directly measured
$v_b$ value for the \ovi\ line. Of course, attributing the entire \ovi\ 
absorption line to the hot gaseous disk represents an extreme
case; the relationship between the \ovi- and \ovii-bearing gases
is still very uncertain (see discussion in \S4). But, if only
part of the \ovi\ absorption is associated with the disk, the lower limit
to the $\gamma$ value will be increased further. 

As a consistency check, we further compare the predicted background emission 
intensity ($I_{\rm SXB}$) from the characterized exponential disk with the 
observed value (\S\ref{sec:sxb}). 
We first revise the thermal emitting gas model {\sl cemekl} (as in XSPEC)
and rename it as {\sl absmkl}, 
\footnote{This model can be obtained from the authors on request.}
which takes $T_0$, $N_H$, $L$,
and $\gamma$ as input parameters and accommodates 
a power law EM of the gas, as presented in Eqs.~\ref{equ:int}-\ref{equ:EMpl}. 
Taking $L=2$ kpc, $\gamma=0.6$, and 
$\log[N_{\rm H}({\rm cm^2})]=19.12$ (Table~\ref{tab:abs}), we 
integrated this model with RASS response matrix to estimate the expected 
hot gas emission intensity in the $\frac{3}{4}$ keV band.
Since the emission measure is mostly sensitive to the plasma
temperature, we include the uncertainty in the mid-plane temperature,
$\log[T_{\rm 0}({\rm K})]=6.44(6.37, 6.52)$ (Table~\ref{tab:abs}). 
We obtain 
$I_{\rm SXB}=1.0(0.4, 1.7)\times10^{-4}\ {\rm counts\ s^{-1}\ armin^{-2}}$,
which, within the errors, is consistent with the observed 
value (\S\ref{sec:sxb}).

\section{Discussion \label{sec:dis}}

\subsection{Comparison with previous works}

We have presented a joint analysis of 
the $z \sim 0$ absorption and emission lines in the X-ray spectra
of Mrk~421 and found that the X-ray absorbing/emitting
gas is consistent with being Galactic, most likely located in a thick
hot gaseous disk with an effective scale length $\sim2$ kpc. This same disk
may also explain much of the \ovi-bearing gas observed in the far-UV \ovi\
line absorption. Here we compare our results and conclusions with 
those from existing studies on the hot gas
along the same sight-line; some of which also give estimates of
the scale length of the X-ray absorbing/emitting gas. All 
existing studies {\sl assume} that the hot gas along the sight-line
is in an isothermal state. We have shown that this assumption
is problematic, at least in the Galactic interpretation. Instead, our analysis
have allowed for nonthermality, which is facilitated by the joint-fit
approach. This approach maximizes the use of the information in the data, 
and propagates the error of each model parameter and takes
care of the correlations between different parameters automatically, and
thus yields more reliable results.
There are, of course, other differences in both the analysis and 
interpretation, sometimes leading to very different conclusions 
in various existing studies.
  
\citet{ras03}, for example,
estimate a scale length $>140$ kpc, based on the averaged \ovii\ and 
\oviii\ line absorptions along the sight-lines toward three AGNs 
(Mrk~421, 3C~273, and PKS~2155-304) and the \ovii\ emission 
line from \citet{mcc02}. This estimate uses an ionization fraction
$f_{\rm i}$ of 0.5 for \ovii\ and an oxygen abundance of 0.3 solar.
The large scale is in contrast to our inferred value $L\sim 2$ kpc
(\S3). The bulk of this discrepancy can be easily accounted for
(see Eq. 3):
oxygen abundance (0.3 solar versus 1 solar), hydrogen density
distribution (uniform versus exponential, causing a factor of 2 difference), 
and the mean hot gas temperature
(6.17 versus 6.35 in logarithm, giving a factor of $\sim 6$ due to the change
of both ionic fraction and emissivity). These differences give a combined 
factor of $\sim 40$; the different adopted atom data may also contribute.
The higher average gas temperature adopted in \citet{ras03} results from 
the inclusion of the 3C~273 sight-line that passes through the inner region of
the Galaxy and is apparently contaminated by the North Polar Spur or
the radio Loop I \citep{sno97}.
Additional difference likely arises from the non-isothermality
of our hot gaseous disk model (\S\ref{sec:NIM}). An absorption line tends
to be produced by gas at lower temperatures than the corresponding emission line of the same transition (e.g., \ovii\ K$\alpha$; compare Fig.~\ref{fig:oxygen}
and Fig.~1 in Yao \& Wang 2005). 

Williams \etal (2005) have also compared the $z\sim0$ X-ray and far-UV 
(\ovi) absorption lines toward Mrk~421, under the isothermal gas assumption. 
They find that the velocity dispersion inferred from the X-ray absorption 
lines is smaller than that of the direct measurement of the \ovi\ line.
As an alternative, an extragalactic
origin of the X-ray-absorbing gas with an inclusion of
the photo-ionization process is explored and is found to be
consistent with the absorption data, which requires a
super-solar abundance ratio of Ne/O. 
In this work, we only consider the 
oxygen absorption lines to avoid the uncertainty in the ratio. We find that, 
while our measured equivalent widths (EWs) for other oxygen lines are 
consistent with those reported by \citet{wil05}, the EW of \ovii\ K$\alpha$
line [11.4(10.20, 12.68) mA] is about $\sim20\%$ higher. Note that our
value is consistent with that obtained by \citet{kaa06}. We believe that
this discrepancy is mainly due to their improper weighted instrumental
response file \footnote{
The response files should be weighted with the corresponding count spectrum 
rather than with the exposure time only, since the flux of the source is 
varying significantly.} for the final co-added spectrum; the higher ($>1$) 
order confusion of the HRC observation (if not properly being taken
into account), the different placement of the continuum, and
the different software packages utilized in the data analysis may also 
contribute.
A higher \ovii\ K$\alpha$/K$\beta$ EW ratio indicates a small saturation
of the \ovii\ K$\alpha$ absorption and therefore requires a large $b_v$,
which then becomes consistent with the value of the \ovi\ line.
Jointly analyzing the absorption and emission data, we have shown here that 
the data can be explained by the Galactic hot gas.

Savage \etal (2005) have explored the {\sl FUSE} spectra of Mrk~421 and two
nearby stars BD~+38$^\circ$2182 ($l,b=182\fdg16, 62\fdg21$)
and HD~93521 ($l, b=183\fdg14, 62\fdg15$)
in great detail. They have examined the relationships 
of the absorption by \ovi\ to those by lowly-ionized species
(e.g., \HI, \cii, \ciii, and \nii) as well as to those by highly-ionized ones
(\ovii\ and \oviii). The observed ``broad'' \ovi\ absorption is divided  into
the two different velocity components, from -140 to 60 km s$^{-1}$ and from
60 to 165 km s$^{-1}$. A good correlation between the positive 
high-velocity \ovi\ component and low ionization gas 
absorptions is found. Interestingly, the high-velocity \ovi\ 
component does not seem to show up in the spectra of the two 
stars. If these sight-lines are not atypical, the positive high velocity 
absorption should then occur at a distance at least 3.5 kpc off the 
Galactic plane. In this case, the absorption cannot arise
from the same location as the \ovii- and \oviii-absorbing gas. 

Some portion of the observed \ovi\ absorptions should be associated with
the \ovii-bearing gas, which we assume to be located in a thick
hot gaseous disk. To further probe this association, 
we reprocessed the {\sl FUSE} observations of 
BD~+38$^\circ$2182, following the standard CalFuse scripts. 
We find that the main \ovi\ absorption along the sigh line
can be characterized as a narrow component with 
$v_b=33(31, 35)\ {\rm km\ s^{-1}}$ and 
log$[N{\rm_{OVI}(cm^{-2})}]=14.20(14.17, 14.22)$. However, because the 
continuum 
placement is very arbitrary (see Zsarg\'o \etal 2003
and Savage \etal 2005 for further discussions), this characterization is
very uncertain. A careful check suggests that up to 30\% of the broad 
\ovi\ line absorption observed in Mrk~421 can be superposed on the narrow
line easily seen in the spectrum of BD~+38$^\circ$2182. This ``common'' broad
component could be associated with the \ovii-bearing gas.
The remaining \ovi\ line absorptions in the sight-lines toward
the star and of Mrk~421 could then arise in the discrete interfaces 
at boundaries of cool clouds, emersed in hot gas \citep{sav05}; the apparent
broadening observed in Mrk~421 may be due to the superposition of different 
velocity components over a large distance. The true absorption and 
emission of the gas should be a convolution of the local temperature and 
density dispersions with the global dependences on the off-plane distance. But
currently there is little observational or theoretical 
guidances as to the specific forms of the dispersions. Our simple
exponential model, hopefully, provides a first-order correction to
the isothermal assumption, or at least demonstrates that the
non-isothermality of the hot gas should be seriously explored.

To examine the influence of the ``common'' component on our results, 
we perform the same joint analysis as in \S\ref{sec:NIM}, but requiring
the hot gaseous disk accounts for only 30\% of the total \ovi\ absorption
detected in the Mkr~421 sight-line.
The constrained parameter ranges are very similar to those illustrated in
Fig.~\ref{fig:Gamma_Tmax} (d)-(f), except for a slightly larger $\gamma$
and a smaller $T_{\rm 0}$ values. 
Clearly, we cannot unambiguously determine the exact fraction of the 
observed \ovi\ absorption that may be associated with X-ray
absorbing gas. But this uncertainty is not expected to 
qualitatively affect our results and conclusions about the overall
non-isothermality and extent of the X-ray-emitting/absorbing gas.

\subsection{Location of the \ovii-bearing gas}
 
The above analysis and results show that the gas that produces the 
X-ray line absorption and emission along the sight-line toward Mrk~421
is consistent with a Galactic hot thick disk interpretation. But based on
the data for this single sight-line, we cannot conclusively determine
the location of the gas. As mentioned above, alternative scenarios
(i.e. Galactic halo and Local Group) of the $z\sim0$ \ovii-bearing gas
have been proposed, allowed 
for by the limited spectral resolution of the X-ray grating observations.
So let us review and discuss the merit and implications of these proposals.

For instance, \citet{nic02} and \citet{wil05} adopted a 
Local Group WHIM interpretation and found a consistent scale length of 
$\sim1$ Mpc of the gas in the sight lines of
PKS 2155-304 and Mrk~421. This consistency is partly a natural outcome
of the low-density assumption, hence the importance of the
photo-ionization by the extragalactic background in
determining the ionic fractions of the gas.
This interpretation, however, requires the WHIM to be either metal-rich 
(e.g., $A_O\ge0.3$ solar) or severely (spherical) asymmetric (i.e., 
a small sky-covering-fraction $f_c$), or both
\citep{wil05, fang06}, because otherwise the derived total baryonic matter 
in the Local Group would be even higher than its gravitational mass
($\sim2\times10^{12}~M_\odot$; Courteau \& van den Bergh 1999).
Recently, \citet{fang06} performed a systematic search for the 
\ovii\ absorption line in AGN spectra from archived 
{\sl Chandra} and {\sl XMM} data and found that the \ovii\ absorption line 
is detected significantly as long as the counting statistic is reasonably
good. They concluded that $f_c$ is $\gsim63\%$
and likely to be unit. This large $f_c$ is  clearly contradictory to
the requirement of the WHIM interpretation that $f_c$ must be $\lsim30\%$ 
even assuming the solar metallicity for the hot gas \citep{fang06}. 
Therefore, the WHIM scenario is very unlikely, if not impossible, to
explain the bulk of the \ovii-bearing gas. 

Alternatively, hot gas is also expected to be present in the Galactic halo
on a scale of  $\sim 100$ kpc. Such a hot gaseous Galactic halo is
predicted in galaxy formation theories. Such gas is heated at 
the so-called virial shocks and through subsequent
gravitational compression (e.g., Birnboim \& Dekel 2003). 
This gas, cooling radiatively, can maintain lasting 
star formation in galactic disks. But simulations tend to 
over-predict the cooled gas content in galactic disks, which 
is the so-called ``over-cooling'' problem (e.g., Navarro \& Steinmetz 1997). 
The inclusion of galactic energy feedback may alleviate the problem, although
this has not been satisfactorily demonstrated. Observationally, 
the existence of a hot gaseous halo around our Galaxy is 
supported by observations of high velocity (HV; 
$|v_{\rm LSR}| \gsim100~{\rm km~s^{-1}}$) clouds (HVCs). Some of these clouds,
especially compact ones, appear to be located far away from the 
Galactic plane, although tight distance constraints are very few.
The hot gaseous halo may be needed to confine such clouds and may even
be their origin through condensation. Furthermore,  HV \ovi\ absorption 
lines are sometimes associated with HVCs and are believed to be produced 
at their interfaces with the hot halo gas \citep{sem03}. 
So far, however, there is little observational evidence for a 
large-scale diffuse X-ray-emitting or -absorbing halo around our Galaxy
or other similar disk galaxies (see Wang 2006 for a review), consistent with
the conclusion reached in the present work. 
A scenario that reconciles these
observations is that much of the hot halo gas is very low in metallicity,
because both the X-ray emission and absorption arise from metal ions.
Such a Galactic hot gaseous halo with low metallicity
may be a natural product of the strong selective radiative
cooling if the accreted gas has a distribution of
metal abundances; the gas with a higher metal abundance cools faster and forms 
denser clouds (probably HVCs), whereas the remaining gas tends 
to have zero or low metallicity, especially in inner regions of 
galactic halos. This scenario explains the low 
X-ray emission/absorption of the hot halo gas and 
further alleviates the over-cooling problem. A detailed modeling 
of the scenario will be presented elsewhere.

We favor the interpretation that the bulk of the $z \sim 0$ \ovii-bearing 
gas represents a hot thick gaseous disk around the Milky Way,
consistent with previous studies as well as the present one. In this case,
the gas represents the hot phase of the ISM and is presumably heated 
primarily by supernova blastwaves. The presence of the gas 
is also traced by \ovi\ absorption. As its ionic fraction peaks at $3 \times
10^5$ K, \ovi\ is believed to present primarily at interfaces between 
cool and 
hot gases and/or in hot gas cooling. The \ovi\ line absorption
is observed ubiquitously at relatively low velocity (LV; $|v_{\rm LSR}| 
\lsim100~{\rm km~s^{-1}}$),
clearly indicating a predominantly Galactic disk origin \citep{wak03}. 
The spatial distribution of the \ovi\ absorption indeed has a 
characteristic vertical scale height 
of $\sim2.3$ kpc around the Galactic plane \citep{sav03} and is thus in a good
agreement with that of the hotter \ovii-bearing gas. Therefore, the
presence of the  hot thick gaseous disk is fully expected and provides
a natural explanation for much of the X-ray/far-UV absorption and emission.

\begin{figure}
  \centerline{
    \includegraphics[width=0.45\textwidth]{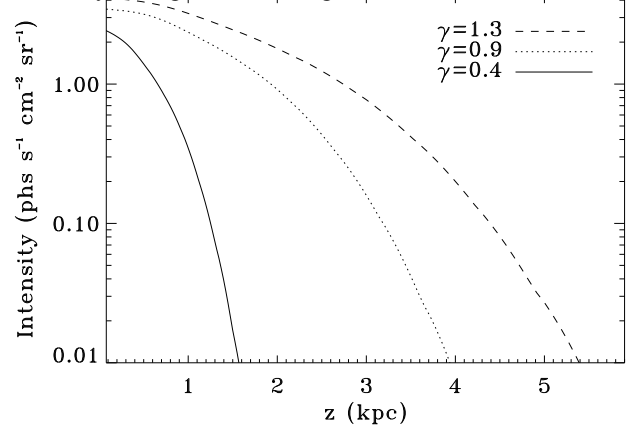}
  }
  \caption{The expected surface intensity distribution in 0.3-1.5 keV
    band as a function of the vertical distance off the Galactic plane. 
    \label{fig:surface}
  }
\end{figure}

The hot thick gaseous disk interpretation is also supported by the X-ray 
emission observations of nearby disk galaxies like our own. Both the intensity
and spatial extent of diffuse X-ray emission appear to be correlated 
with star formation in galactic disks. Typically the emission is
confined to galactic coronae around the disks with vertical extent no
more than a few kpc (e.g., \citet{tul06}, Wang \etal 2003). For ease
of comparison with these studies, we present  in Fig.~\ref{fig:surface}
the emission intensity distribution as a function of the vertical distance 
away from the Galactic plane with the best-fit parameters of the
hot thick gaseous disk (\S\ref{sec:NIM}). Because the $\gamma$ is not well 
constrained, the distribution for several specific 
$\gamma$ values are included. For a very broad range of $\gamma$, the 
emission is essentially concentrated within a few kpc off-plane distance.
This concentration is consistent with those findings 
in the nearby disk galaxies \citep{wang03,tul06},
suggesting a common behavior and a common nature of hot gas
in these galaxies.

To summarize, we have carried out a systematic investigation of the available
X-ray \ovii\ and \oviii\ absorption/emission
tracers of Galactic diffuse hot gas in the field of Mrk~421. 
A joint analysis of the data, complemented by the {\sl FUSE} \ovi\
measurement, enables us for the first time to self-consistently and
simultaneously characterize the non-isothermality and vertical
scale height of the hot gas in the sight-line toward Mrk~421.
We have established that the X-ray-emitting/absorbing
hot gas cannot be in an isothermal state. But the X-ray data can be
explained with a hot gaseous disk model, which is meant
to provide a simple, yet physically
plausible characterization of the global temperature and density distributions
of the hot gas. Indeed, our inferred scale height of the hot gaseous disk 
and the gas density at the Galactic plane are
consistent with our previous estimations based on the Galactic
latitude dependence of Galactic \neix\ line absorptions 
observed \citep{yao05}.

\acknowledgments
We thank Dan McCammon and Claude Canizares for useful discussions and 
the second referee 
for insightful comments and constructive suggestions, which helped to improve
our presentation of the paper. 
We are also grateful to Michael Nowak for investigating the possible 
differences in measuring absorption line equivalent width in ISIS, XSPEC, 
and Sherpa.
This work is supported by NASA through the 
Smithsonian Astrophysical Observatory (SAO) contract SV3-73016 to 
MIT for support of the Chandra X-Ray Center, which is operated by the 
SAO for and on behalf of NASA under contract NAS 08-03060. 
Support from the {\sl Chandra} archival research grant AR6-7023X is also 
acknowledged.

\clearpage

\thispagestyle{empty}

\clearpage

\end{document}